\documentclass[preprintnumbers,nofootinbib,aps,11pt, showpacs]{revtex4}

\usepackage{bm,amsmath,amssymb}
\usepackage[dvips]{graphicx}
\usepackage{color}

\begin{document}

\title{{\Large
Integrability of Particle System around a Ring Source\\ 
as the Newtonian Limit of a Black Ring}}

\hfill{{\small OCU-PHYS-415, AP-GR-117, KEK-TH-1783, KEK-Cosmo-159}}


\author{Takahisa Igata$^1$}
\email{igata@kwansei.ac.jp}
\author{Hideki Ishihara$^2$}
\email{ishihara@sci.osaka-cu.ac.jp}
\author{Hirotaka Yoshino$^3$}
\email{hyoshino@post.kek.jp}

\affiliation{%
$^1$Graduate School of Science and Technology, 
Kwansei Gakuin University, 
Sanda, Hyogo, 669-1337, Japan\\
$^2$Department of Mathematics and Physics,
Graduate School of Science, Osaka City University,
Osaka, 558-8585, Japan\\
$^3$Theory Center, Institute of Particles and Nuclear Studies, 
KEK, Tsukuba, Ibaraki, 305-0801, Japan
}

\begin{abstract}
The geodesic equation in the five-dimensional singly rotating black ring 
is non-integrable unlike the case of the Myers-Perry black hole.  
In the Newtonian limit of the black ring, 
its geodesic equation agrees with the equation of motion of 
a particle in the Newtonian potential due to 
a homogeneous ring gravitational source. 
In this paper, we show that the Newtonian equation of motion 
allows the separation of variables in the spheroidal coordinates, 
providing an non-trivial constant of motion quadratic in momenta. 
This shows that the Newtonian limit of a black ring 
recovers the symmetry of its geodesic system, 
and the geodesic chaos is caused by relativistic effects. 
\end{abstract}

\maketitle

\section{Introduction}
\label{sec:1}

The Kerr geometry in four-dimensional general relativity 
possesses remarkable properties, 
such as black hole thermodynamics and the uniqueness of
stationary, axisymmetric, asymptotically flat vacuum solution of 
the Einstein equation with a regular event horizon 
(see, for example,~\cite{Wald:1984rg}). 
One of the most important features is 
the integrability of its geodesic equation. 
This was proven by showing the separability 
of the Hamilton-Jacobi equation for the geodesic system 
in the Boyer-Lindquist coordinates and 
the existence of an additional constant of motion, 
called the Carter constant~\cite{Carter1968}. 
This non-trivial constant is given by a second-rank Killing tensor 
in the Kerr geometry~\cite{Walker1970}, 
which is a generalization of Killing vectors to symmetric tensors. 
Consequently, the separation of variables 
has great advantages in studying 
freely falling particle motion around a Kerr black hole.

For the last few decades, 
many efforts are devoted to studying the higher-dimensional 
black hole solutions of the Einstein equation 
and their physical properties highly motivated 
by superstring theories and the scenarios 
with large extra dimensions~(see, for example,~\cite{LivingReview}). 
In those works, it turned out that 
the stationary rotating black holes in higher dimensions 
do not inherit all of the properties in the four-dimensional Kerr geometry. 
For example, the Myers-Perry geometries~\cite{Myers:1986un}, 
which describe black holes with spherical horizon topology, 
have no uniqueness in stationary rotating 
black hole solutions in higher dimensions. 
However, the integrability of the geodesic equation remains valid 
in the Myers-Perry geometry. 
In five dimensions, 
the separation of variables occurs in the Hamilton-Jacobi equation 
for geodesic system~\cite{Frolov:2003en}. 
After these findings, 
the geodesic equation in the most general known vacuum black hole solution 
with a horizon of spherical topology 
is found to be integrable 
in all dimensions (see, for a review,~\cite{Yasui:2011pr}).

In five-dimensional spacetimes, 
there also exist the black hole solutions that have unusual horizon topology, 
called the black ring solutions~\cite{Emparan:2001wn, Pomeransky:2006bd, 
Mishima:2005id, Elvang:2007rd, Iguchi:2007is, Evslin:2007fv, Izumi:2007qx, Elvang:2007hs}. 
The simplest one is the Emparan-Reall black ring geometry~\cite{Emparan:2001wn}, 
which possesses ${\mathrm S}^2\times {\mathrm S}^1$ horizon topology 
and rotates on its symmetric axis. 
Unlike the case of the Myers-Perry geometry, 
the Hamilton-Jacobi equation for the geodesic system 
in the black ring geometry is not separable 
at least in the ring coordinates except for special cases of 
geodesics~\cite{Hoskisson:2007zk, Durkee:2008an}.\footnote{Other discussions 
on geodesics in the black ring geometry can be seen 
in \cite{Nozawa:2005eu, Igata:2010ye, Igata:2013be, Grunau:2012ai, Grunau:2012ri}.} 
Although non-separability of the Hamilton-Jacobi equation is a necessary condition 
for the non-integrability of the geodesic equation,  
there exists a strong indication for 
the non-integrability of the geodesic equation. 
Indeed, the existence of chaotic behavior of geodesics 
in the black ring geometry 
was demonstrated by the Poincar\'e map method~\cite{Igata:2010cd}. 
Recently, the absence of the Killing-Yano and conformal Killing-Yano tensors 
in the black ring geometry has been rigorously proved in \cite{Houri:2014hma}. 
This result indicates that the black ring geometry is 
less symmetric compared to the Myers-Perry geometry.
However, since the existence of constants of geodesic motion 
is related to the Killing tensor, the mathematical proof of non-integrability 
of the geodesic system in black ring spacetimes still remains an open problem.

Chaos is universal phenomena in various nonlinear systems, 
and the appearance of geodesic chaos also is 
one of the important issues in general relativity. 
In the Majumdar-Papapetrou geometry~\cite{Majumdar1947, Papapetrou1947} 
that represents two-fixed extremal black holes 
there exist geodesic chaos~\cite{Contopoulos:1990, Contopoulos:1991, 
Dettmann:1994dj, Dettmann:1995ex, Yurtsever:1994yb, Sota:1995ms}. 
The results in \cite{Hanan:2006uf} suggest that its geodesic system is chaotic 
in the higher-dimensional Majumdar-Papapetrou geometry as well. 
In the Newtonian limit of four-dimensional two-fixed black holes, 
the geodesic system reduces to Newtonian particle system 
around two-fixed centers, which is known 
as Euler's three-body problem~\cite{Euler:1760}. 
This reduced system is integrable, 
because there exists an additional constant that is quadratic in momentum. 
Therefore, the geodesic chaos in the Majumdar-Papapetrou geometry 
is caused by relativistic effects.

Thus, a natural question arises 
in the study of geodesic chaos in the black ring geometry:
whether or not the chaos is caused by relativistic effects. 
The purpose of this paper is to demonstrate that 
the geodesic chaos does not appear in the weak-field approximation, 
and thus it is generated by strong gravitational effects. 
In order to show this, we show the integrability of 
the geodesic equation in the black ring geometry 
in the Newtonian limit.

This paper is organized as follows. 
In the following section, 
the Newtonian limit of the geodesic equation 
in the singly rotating black ring geometry is demonstrated. 
In Sec.~\ref{sec:3}, 
the Newtonian potential obtained in Sec.~\ref{sec:2} is shown to agree with 
the one generated by a homogeneous ring gravitational source. 
In Sec.~\ref{sec:4}, 
the Newtonian equation of particle motion 
around the ring source is demonstrated to allow 
the separation of variables in the spheroidal coordinates, 
and provide an additional constant of motion that is quadratic in momentum. 
Finally, the implication of our results is discussed in Sec.~\ref{sec:5}.

~~

\section{Newtonian limit of the geodesic equation in the singly rotating black ring geometry}
\label{sec:2}

In this section, we derive a Newtonian potential 
from the Newtonian limit of the geodesic equation in the thin black ring geometry. 
The Newtonian limit of the geodesic equation 
requires slow motion and  weak gravitational field limit. 
Let $X^\mu(\tau)$ be the world line of a freely falling particle with a proper time $\tau$.
For the motion much slower than the speed of light, i.e., in the slow motion limit, 
the four velocity $dX^\mu/d\tau$ is approximately one 
for the time component and zero for the spatial components. 
In the weak field limit, a metric $g_{\mu\nu}$ is decomposed as 
\begin{align}
g_{\mu\nu}=\eta_{\mu\nu}+h_{\mu\nu}
\label{eq:h}
\end{align}
in global inertial coordinates, 
where $\eta_{\mu\nu}$ denotes the flat metric and $|h_{\mu\nu}|\ll 1$. 
Hence the geodesic equation in the Newtonian limit becomes
\begin{align}
\frac{d^2X^\mu}{dt^2}+\Gamma^\mu{}_{00}=0,
\label{eq:geodesic equation}
\end{align}
where the second term of the Christoffel symbol 
in the linear approximation of $h_{\mu\nu}$ is given by
\begin{align}
\Gamma^\mu{}_{00}= -\frac12 \,\eta^{\mu\nu}\partial_\nu h_{00}.
\end{align}
Finally, 
\eqref{eq:geodesic equation} reduces to the Newtonian equation of motion of a particle
subjected to the gravitational force due to the Newtonian potential
that is defined as 
\begin{align}
\phi(\mbox{\boldmath $r$})=-\frac12\, h_{00}. 
\label{eq:phi}
\end{align}
Note that the time derivative of $\phi$ is neglected, 
because we consider a stationary spacetime.

Let us determine the explicit form of $\phi(\mbox{\boldmath $r$})$ 
in the case of the black ring geometry. 
The five-dimensional singly rotating black ring metric 
in the ring coordinates is of the form
\begin{align}
ds^2=& - \frac{F(y)}{F(x)}
\left(dt-CR\,\frac{1+y}{F(y)}\, d\psi\right)^2 
+\frac{R^2F(x)}{(x-y)^2}
\left(- \frac{G(y)}{F(y)}\,d\psi^2- \frac{dy^2}{G(y)}
+ \frac{dx^2}{G(x)}+ \frac{G(x)}{F(x)}\,d\phi^2
\right),
\label{eq:metric}
\end{align}
with
\begin{align}
&F(\chi)=1+\lambda\,\chi,
\\
&G(\chi)=\left(1-\chi^2\right)\left(1 + \nu\, \chi\right),
\\
&C=\sqrt{\lambda \left(\lambda-\nu\right) \frac{1+\lambda}{1-\lambda}}, 
\end{align}
where $R$, $\lambda$, and $\nu$ are constants 
and characterize the ring radius, 
rotational velocity, and thickness of the black ring horizon, respectively. 
From the regularity of the solutions, $\lambda$ must be 
\begin{align}
\lambda=\frac{2\,\nu}{1+\nu^2},
\end{align} 
and then the regular black ring solutions have two-free parameters $R$ and $\nu$. 
Note that 
absence of naked singularities requires the thickness parameters to be 
$0<\nu<1$.

We introduce the polar coordinates given by 
\begin{align}
&x=\frac{R^2-\zeta^2-\rho^2}{\Sigma},
\\
&y=-\frac{R^2+\zeta^2+\rho^2}{\Sigma}, 
\end{align}
and 
\begin{align}
\Sigma=\sqrt{\big(\left(\zeta+ R\right)^2+\rho^2\,\big)
\big(\left(\zeta- R\right)^2+\rho^2\,\big)},
\label{eq:Sigma}
\end{align}
where the flat limit of the metric in these coordinates is of the form
\begin{align}
ds^2
=-dt^2+d\zeta^2+\zeta^2\, d\psi^2+d\rho^2+\rho^2 \,d\phi^2. 
\end{align}
The time-time component of the metric~\eqref{eq:metric} in the new coordinates 
is given by
\begin{align}
g_{00}=-\frac{\Sigma-\lambda\,\big(R^2+\zeta^2+\rho^2\big)}{\Sigma+\lambda\, \big(R^2-\zeta^2-\rho^2\big)}. 
\label{eq:g_00}
\end{align}
From \eqref{eq:h} and \eqref{eq:g_00}, 
the time-time component of $h_{\mu\nu}$ is given by
\begin{align}
h_{00}=\frac{2\lambda R^2}{\Sigma+\lambda \left(R^2-\zeta^2-\rho^2\right)}. 
\end{align}
The leading order of the expansion in $\nu$ takes the form
\begin{align}
h_{00}=\frac{2\lambda R^2}{\Sigma}.
\label{eq:h00}
\end{align}
Note that the other components of $h_{\mu\nu}$ defined 
in \eqref{eq:h} must be much smaller than unity in the case $\nu \ll 1$, 
because the metric \eqref{eq:metric} goes to flat
in the limit $\nu \to 0$. 
Finally, Eqs.~\eqref{eq:phi} and \eqref{eq:h00} lead
\begin{align}
\phi(\mbox{\boldmath $r$})
=-\frac{GM}{2\,\Sigma},
\label{eq:potential}
\end{align}
where 
\begin{align}
M=\frac{4\,\nu R^2}{G},
\label{eq:ADMmass}
\end{align} 
which is the Arnowitt-Deser-Misner mass 
of the singly rotating black ring geometry~\cite{Emparan:2006mm, Tanabe:2010ax}
linearized in $\nu$.\footnote{
The choice of the gravitational coupling constant here gives the Einstein equation in the form
\begin{align}
R_{\mu\nu}-\frac12 R\, g_{\mu\nu}= \alpha \,T_{\mu\nu}
\label{eq:Einstein equation}
\end{align}
with
\begin{align}
\alpha=\frac{D-2}{D-3}\,S_{D-2}\,G_D, 
\label{eq:alpha}
\end{align}
where $S_{D-2}$ and $G_D$ denote the area of $(D-2)$-sphere and 
gravitational constant in $D$ dimensions, respectively, 
and $D=5$ in this case.}
Hence, the Newtonian limit of the geodesic equation 
in the singly rotating thin black ring geometry reduces to 
the Newtonian equation of motion of a particle subjected to 
gravitational force due to the Newtonian potential~\eqref{eq:potential}. 
Since the topology of the curvature singularity in the black ring geometry 
is ring shape and the horizon in this case is very thin, 
it is expected that the matter distribution producing \eqref{eq:potential} is a ring source.

~~

\section{Newtonian potential due to a gravitational ring source}
\label{sec:3}

In this section we provide the explicit form of the Newtonian potential 
due to a homogeneous ring source placed 
on four-dimensional flat space by integrating the basic equation 
of higher-dimensional Newton gravity. 
As will be seen in the last of this section, 
the Newtonian potential takes the same form of 
the potential obtained in the previous section.

\begin{figure}[t]
 \includegraphics[width=7.0cm,clip]{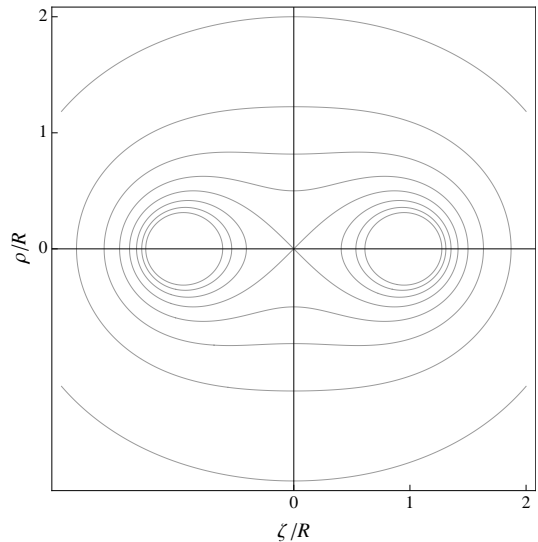}
 \caption{
Equipotential surfaces produced by a ring source with radius $R$ in the $\zeta$-$\rho$ plane in the case $G M/R^2=1$. The contours show the Cassini ovals. 
}
 \label{fig:equipotential surfaces}
\bigskip
\end{figure}

By analogy with the configuration of the curvature singularity of the black ring geometry,
let us place a homogeneous ring source on a circle of radius $R$ 
from the coordinate origin in the two-dimensional plane with $\rho=0$. 
Then the mass density is of the form 
\begin{align}
\sigma(\mbox{\boldmath $r$})=\frac{M}{(2\pi)^2\,\zeta\, \rho} \,\delta(\zeta-R)\,\delta(\rho),
\label{eq:density}
\end{align}
where $M$ denotes the total mass of the ring source in this section 
and $\delta(\cdot)$ is the delta function. 
The basic equation to determine the form of gravitational potential 
is the Newtonian gravitational field equation in higher dimensions 
\begin{align}
\nabla^2 \phi(\mbox{\boldmath $r$})=S_{3}\, G\, \sigma(\mbox{\boldmath $r$}),
\label{eq:Poisson's equation}
\end{align}
where $\nabla^2$ denotes Laplacian of the flat space, 
$G$ is the gravitational constant in five dimensions. 
Note that this equation is consistent with 
the Newtonian limit of the Einstein equation given in 
\eqref{eq:Einstein equation} and \eqref{eq:alpha} 
up to the numerical factor. 
The solution to \eqref{eq:Poisson's equation} is written in the form
\begin{align}
\phi(\mbox{\boldmath $r$})=-\frac{G}{2} \int_{{\mathbb R}^4}\, \frac{\sigma(\mbox{\boldmath $r$}')}{|\mbox{\boldmath $r$}-\mbox{\boldmath $r$}'|^2}\,dV(\mbox{\boldmath $r$}'),
\label{eq:sol}
\end{align}
where $dV$ is the volume element on ${\mathbb R}^4$.
The integration of \eqref{eq:sol} with the matter distribution~\eqref{eq:density} 
provides the explicit form of the Newtonian potential due to the ring source as
\begin{align}
\phi(\mbox{\boldmath $r$})
=-\frac{GM}{2\,\Sigma}. 
\label{eq:potential2}
\end{align}
Thus \eqref{eq:potential2} agrees with \eqref{eq:potential} 
under the identification of the total mass $M$ of the homogeneous ring source 
as the Arnowitt-Deser-Misner mass of the black ring~\eqref{eq:ADMmass}. 
Thus, it is shown that the gravitational field of the black ring 
in the Newtonian limit reduces to the Newtonian potential produced 
by a homogeneous ring source. 
The equipotential surfaces of \eqref{eq:potential2}, or equivalently, of \eqref{eq:potential} 
are shown in Fig.~\ref{fig:equipotential surfaces}. 

~~

\section{Separability of the Hamilton-Jacobi equation}
\label{sec:4}

In this section, let us consider the Hamiltonian system of particle motion 
in the gravitational field due to a homogeneous ring source. 
In order to show the integrability of this system, 
the separability of the Hamilton-Jacobi equation 
is demonstrated in the spheroidal coordinates. 
As a result, we obtain the expression for a constant of motion, 
which is analogous to the Carter constant in the Kerr geometry, 
in addition to the energy and the two angular momenta.

The Hamiltonian of the particle system in the gravitational field \eqref{eq:potential2} 
is given by\footnote{
Particles described by \eqref{eq:H} 
have stable bound orbits by a balance between 
centrifugal force and gravitational force due to the ring source, 
which is consistent to the existence of stable bound orbits 
at a distant region from a black ring with $\nu <1/3$ 
as demonstrated in \cite{Igata:2010ye}.} 
\begin{align}
H
=\frac{1}{2m}\bigg(\,p_\zeta^2+\frac{p_\psi^2}{\zeta^2}
+p_\rho^2+\frac{p_\phi^2}{\rho^2}\,\bigg)
-\frac{GMm}{2\,r_+r_-},
\label{eq:H}
\end{align}
where $m$ is the mass of a particle,  
$p_i$ is the momentum of a particle canonically conjugate to the polar coordinates, 
and $r_\pm$ are defined by
\begin{align}
r_\pm=\sqrt{(\,\zeta\pm R\,)^2+\rho^2}.
\end{align}
Note that the product of $r_+$ and $r_-$ agrees with \eqref{eq:Sigma}, 
\begin{align}
\Sigma=r_+ r_-.
\end{align}
The Hamiltonian shows that the Hamilton-Jacobi equation of this system does not 
lead to the separation of variables in the polar coordinates.

Here we introduce the spheroidal coordinates $(\xi, \eta)$\footnote{
The transformation also is written in the simple form
\begin{align}
&r_\pm=R\left(\xi\pm \eta\right).
\end{align}
} 
defined by
\begin{align}
&\zeta=R\,\xi\, \eta,
\label{eq:zeta}\\
&\rho=R\sqrt{\left(\xi^2-1\right)\left(1-\eta^2\right)}. 
\label{eq:rho}
\end{align}
\begin{figure}[t]
 \begin{center}
 \includegraphics[width=7cm,clip]{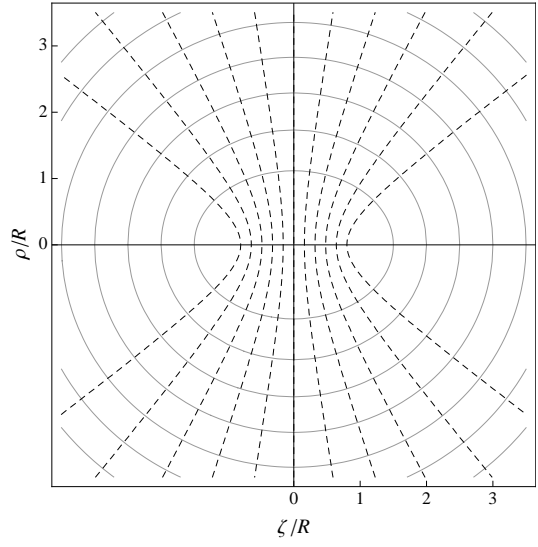}
 \end{center}
 \caption{
Relation between the polar coordinates $(\zeta, \rho)$ 
and the spheroidal coordinates $(\xi, \eta)$. 
The solid lines shows constant $\xi$, which are confocal ellipsoids, 
and the dashed lines shows constant $\eta$, which are confocal hyperbolas, 
where the focal point is located at $(\zeta, \rho)=(R,0)$. 
}
 \label{fig:spheroidal coordinates}
\bigskip
\end{figure}
The contours for $\xi$ and $\eta$ in $(\zeta, \rho)$-plane 
are shown in Fig.~\ref{fig:spheroidal coordinates}. 
In the spheroidal coordinates, 
the four-dimensional flat metric is of the form
\begin{align}
ds^2=R^2\left(\xi^2-\eta^2\right)\bigg(\, \frac{d\xi^2}{\xi^2-1}+\frac{d\eta^2}{1-\eta^2}\,\bigg)
+R^2\,\xi^2\,\eta^2 \,d\psi^2+R^2\left(\xi^2-1\right)\left(1-\eta^2\right)d\phi^2.
\end{align}
The Hamiltonian \eqref{eq:H} is given in the spheroidal coordinates as
\begin{align}
H=&\,\frac{1}{2mR^2\left(\xi^2-\eta^2\right)}\bigg[
\left(\xi^2-1\right)\,p_\xi^2
+\left(1-\eta^2\right)\,p_\eta^2
\cr
&+\bigg(\,\frac{1}{\eta^2}-\frac{1}{\xi^2}\,\bigg)\,p_\psi^2
+\bigg(\,\frac{1}{\xi^2-1}+\frac{1}{1-\eta^2}\,\bigg)\,p_\phi^2\,\bigg]
-\frac{GMm}{2R^2(\xi^2-\eta^2)},
\end{align}
From this form of $H$, the Hamilton-Jacobi equation is given by
\begin{align}
&\frac{1}{2mR^2\left(\xi^2-\eta^2\right)}\bigg[
\left(\xi^2-1\right)\bigg(\frac{\partial S}{\partial \xi}\bigg)^2
+\left(1-\eta^2\right)\bigg(\frac{\partial S}{\partial \eta}\bigg)^2
\cr
&+\left(\frac{1}{\eta^2}-\frac{1}{\xi^2}\right)
\bigg(\frac{\partial S}{\partial \psi}\bigg)^2
+\left(\frac{1}{\xi^2-1}+\frac{1}{1-\eta^2}\right)\bigg(\frac{\partial S}{\partial \phi}\bigg)^2-GMm^2\,\bigg]-\frac{\partial S}{\partial t}=0,
\label{eq:HJeq}
\end{align}
where $S$ is Hamilton's principal function. 
To find a complete solution to the Hamilton-Jacobi equation, 
$S$ is assumed to have 
the completely separated form
\begin{align}
S=p_\psi\, \psi+p_\phi \,\phi +S_\xi(\xi)+S_\eta(\eta)-E \,t,
\label{eq:S}
\end{align}
where $p_\psi$, $p_\phi$, and $E$ are conjugate momenta 
to $\psi$, $\phi$, and $t$, respectively, 
which are constants of motion, 
and $S_\xi(\xi)$ and $S_\eta(\eta)$ are the 
functions of $\xi$ and $\eta$, respectively, to be determined later. 
Substitution of \eqref{eq:S} into \eqref{eq:HJeq} shows 
that the Hamilton-Jacobi equation allows 
the separation of variables such that
\begin{align}
-P(\xi)=Q(\eta)=C,
\end{align}
where
\begin{align}
&P(\xi)=\left(\xi^2-1\right)\left(\frac{d S_\xi}{d\xi}\right)^2
-\frac{p_\psi^2}{\xi^2}+\frac{p_\phi^2}{\xi^2-1}-2mER^2\xi^2-\frac{GMm^2}{2},
\\
&Q(\eta)=\left(1-\eta^2\right)\left(\frac{d S_\eta}{d\eta}\right)^2
+\frac{p_\psi^2}{\eta^2}+\frac{p_\phi^2}{1-\eta^2}+2mE R^2 \eta^2-\frac{GMm^2}{2},
\end{align}
and $C$ is the constant of the separation. 
This additional constant of motion $C$ is analogous to 
the Carter constant for the geodesic system in the Kerr geometry.

As a result, the Newtonian particle system~\eqref{eq:H} 
is separable in the spheroidal coordinates. 
In addition, the set of the constants of motion, 
$H$, $p_\psi$, $p_\phi$, and $C$, commute with 
each other under the Poisson bracket, which means that 
those constants are independent and 
this particle system is integrable in Liouville's sense. 
Hence, it is concluded that 
chaotic orbits of a particle in a five-dimensional black ring is caused by relativistic effect.

For physical understanding, let us rewrite $C$ in the polar coordinates as
\begin{align}
C
&=L^2+R^2\,\left(p_\zeta^2+\frac{p_\psi^2}{\zeta^2}\right)+m^2f, 
\label{eq:C}
\end{align}
where
\begin{align}
&L^2
=\left(\zeta \,p_\rho-\rho\, p_\zeta\right)^2+\left(\zeta^2+\rho^2\right)
\left(\frac{p_\psi^2}{\zeta^2}+\frac{p_\phi^2}{\rho^2}\right),
\\
&f=-\frac{GM}{4}\, \frac{r_+^2+r_-^2}{r_+r_-}. 
\label{eq:f}
\end{align}
Note that $L^2$ is the squared total angular momentum. Therefore, 
the constant $C$ is essentially interpreted as the squared total angular momentum, 
because \eqref{eq:potential} in the limit of a point source, $R\to 0$, goes to central force potential, and $C$ reduces to $L^2$ except for constant shift. 
Note that the quadratic terms in the momenta in \eqref{eq:C}, $C^{ij}p_i\,p_j$,
are written by the Killing tensor in the four-dimensional flat space, 
\begin{align}
C^{ij}
=\left(\rho^2+R^2\right)\partial_\zeta^i \,\partial_\zeta^j
+\zeta^2\,\partial_\rho^i \,\partial_\rho^j
-2\,\zeta\,\rho\,\partial_\zeta^{(i}\,\partial_\rho^{j)}+\frac{\zeta^2+\rho^2+R^2}{\zeta^2}\,\partial_\psi^i \,\partial_\psi^j+\frac{\zeta^2+\rho^2}{\rho^2}\,\partial_\phi^i \,\partial_\phi^j.
\label{eq:Cij}
\end{align}
Note that $C^{ij}$ is reducible, i.e., 
a linear combination of symmetric tensor product of the Killing vectors 
in the four-dimensional flat space, 
because the background is maximally symmetric, and accordingly, 
there exist only reducible second-rank Killing tensors in contrast to 
the Kerr geometry. 
The pair of \eqref{eq:Cij} and \eqref{eq:f} solves the Killing hierarchy equation given in \cite{Igata:2010ny}.

~~

\section{Discussion}
\label{sec:5}

In this paper, the integrability of the geodesic equation has been investigated
in the singly rotating black ring geometry in the Newtonian limit. 
The geodesic equation in the thin black ring geometry reduces to 
the Newtonian equation of motion of 
a particle moving in the gravitational potential generated by 
a homogeneous ring source. 
This is consistent with the fact that 
the curvature singularity of the black ring has S$^1$ topology and 
the event horizon is thin in this case. 
The main result is that the Newtonian equation of particle motion allows 
the separation of variables in the spheroidal coordinates, and then, 
there exists an additional constant of motion quadratic 
in momentum analogous to the Carter constant in the Kerr geometry. 
This means that the Newtonian system of particle motion 
around a ring source is integrable. 
Hence, it is concluded that the appearance of geodesic chaos 
in the black ring geometry is caused by relativistic effects.

Our result implies the existence of an approximate constant for the geodesic system and 
an approximate Killing tensor in the thin black ring geometry. 
The approximate Killing tensor must be useful in the study 
for the physics in the black ring geometry, e.g., field dynamics and perturbation. 
Clarifying the condition for break down of 
the integrability of the geodesic system is one of the interesting 
issues for future work. 

~~

\subsection*{Acknowledgements}
This work is partially supported by a Research Grant from the Tokyo Institute of Technology Foundation (T.I.), Grant-in-Aid for Scientific Research No.~24540282 (H.I.), and Grant-in Aid for Scientific Research (A) No. 26247042 (H.Y.).

~~

\end{document}